\documentclass[conference]{IEEEtran}
\IEEEoverridecommandlockouts
\usepackage[dvips]{graphicx}
\usepackage{subcaption}
\usepackage{epsfig}
\usepackage{amsmath}
\usepackage{epsfig}
\usepackage{graphicx}
\usepackage{makecell}
\usepackage{textcomp}
\usepackage{xcolor}
\usepackage{subcaption}
\usepackage{array}
\usepackage{amssymb}
\usepackage{color}
\usepackage{lipsum}
\usepackage{algorithm}
\usepackage{algpseudocode}

\usepackage{geometry}
\title{Optimization of MIMO STEEP for Secure Communications over MIMOME Channels}
\author{\IEEEauthorblockN{Dinglin Yu and Yingbo Hua}
\IEEEauthorblockA{\textit{Department of Electrical and Computer Engineering} \\
\textit{University of California at Riverside}\\
Riverside, CA 92521, USA \\
dinglin.yu@email.ucr.edu and yhua@ucr.edu}
}

\begin{document}

\maketitle
\begin{abstract}
A transmission scheme for secure communications, called secret-message transmission by echoing encrypted probes (STEEP), can deliver a positive secrecy rate, even when eavesdropping channels are much stronger than that between users, subject to sufficient asymmetric power allocations from users. This paper considers an optimization problem for STEEP applied to MIMO (or MIMOME) channels between multi-antenna users and multi-antenna Eve. We reveal for the first time a significant increase of the achievable secrecy rate of MIMO STEEP using optimized precoders.
\end{abstract}


\section{Introduction}

Secure communication without a pre-existing secure channel or a pre-shared secret key has been a long-standing problem, which has been handled by two primary approaches. The first is based on some discrete hard-to-invert function for two (or more) parties to generate their own public keys from their private keys, where the public keys are then exchanged to generate a shared secret key to be used for subsequent secure communications. Any infrastructure or standardization based on this approach is however vulnerable to future attacks by increasingly powerful computers. The other approach, also known as physical layer security (PLS), is based on information-theoretical security, which is not breakable by any computing power. This paper is concerned with the second approach.

There are numerous works on PLS in the literature, and they can be classified into two groups: one is based on single-trip wiretap channel (ST-WTC) schemes where a secret-message is transmitted from one user to another via a single-trip coding scheme, and the other is based on secret-key generation (SKG) where correlated data sets collected at two users are used to generate a shared secret key.

It is known \cite{KhistiWornell2010} and \cite{OggierHassibi2011} that, when using ST-WTC schemes, the achievable secrecy rate (in bits per channel use) from Bob to Alice  is zero whenever Eve's receive channel capacity from Bob cannot be made larger than Alice's receive channel capacity from Bob. Such a situation often happens when Eve's distance to Bob is shorter than Alice's and at the same time Eve's number of antennas is no less than Bob's.

On the other hand, it is also known \cite{RennaBloch2013} and \cite{HuaMaksud2024} that the secret-key capacity (in bits per probing channel use) for SKG from channel probing and information reconciliation does not have the above mentioned limitation. But secret-key capacity does not take into account the channel uses needed for information reconciliation (IR). Many proposed methods in the literature for IR require a substantial amount of additional channel uses.

An early attempt to take advantage of both SKG and WTC was shown in \cite{Hayashi2020}, where the authors considered binary signaling over SISO channels.  Recently, a much more general fusion of SKG and WTC has resulted in what is called secret-message transmission by echoing encrypted probes (STEEP) \cite{Hua2023}, \cite{Hua2025}, \cite{HuaICC2024}, \cite{HuaLANMAN2024}, \cite{Rahman2024}, \cite{Rafique2025}. Unlike any ST-WTC scheme, STEEP utilizes a round-trip (RT) coding scheme inspired by  an encryption lemma shown in \cite{Bloch2011} and \cite{Gamal2011}, which allows STEEP to yield a positive secrecy rate by an asymmetric power control under virtually any channel conditions. In other words, via a RT coding scheme, STEEP creates a virtual  WTC system where Eve's effective receive channel is always disadvantaged even if her original or raw receive channels are stronger than users'.
The above property holds in general, including the case where STEEP is applied to MIMO channels.

However, the current MIMO version of STEEP as shown in \cite{Hua2025} is not optimized, which does not exploit any possible knowledge of Eve's channel matrices. The aim of this paper is to show how to optimize the  precoding matrices for MIMO STEEP, assuming that all channel matrices are given. This will allow us to see the secrecy rate gap between the optimized MIMO STEEP and a baseline MIMO STEEP.

It is important to note that for the MIMO ST-WTC scheme as studied in \cite{KhistiWornell2010} and \cite{OggierHassibi2011}, there are many prior works on optimization of the transmit covariance matrix to maximize its secrecy rate. See \cite{Shi2011,Li2013,Mukherjee2022} and the references therein. But the secrecy rate expression for the MIMO STEEP is very much different from that for the MIMO ST-WTC. The former is more complicated than the latter. What is similar here to prior works is that we also apply such principles as alternating optimization (AO), weighted minimum mean squared error (WMMSE), successive convex approximation (SCA), and projected gradient ascent (PGA).

\section{Channel Model and MIMO STEEP}\label{sec:model}
Consider a network of MIMO channels among two legitimate users (Alice and Bob) and an eavesdropper (Eve).
Alice and Bob are equipped with $n_1$ and $n_2$ antennas, respectively. Eve possesses $l_1$ antennas for intercepting in the probing phase, and $l_2$ antennas for intercepting in the echoing phase.
We adopt a block-fading narrowband channel model where all channel matrices are constant within each channel coherence period of $T_c$ second while all signals and noises at different sampling/symbol times within each $T_c$ are modelled as independent and identically distributed (i.i.d.).

Let $\mathbf{x}_1(k)$ be the $k$th symbol transmitted by Alice in phase 1 (probing phase), and
the corresponding signals received by Bob and Eve are respectively $\mathbf{y}_2(k)$ and $\mathbf{z}_1(k)$:
\begin{equation}\label{phase1}
    \begin{aligned}
        \mathbf{y}_2(k)&=\mathbf{H}_2\mathbf{x}_1(k)+\mathbf{w}_2(k),\\
        \mathbf{z}_1(k)&=\mathbf{G}_1\mathbf{x}_1(k)+\mathbf{v}_1(k).
    \end{aligned}
\end{equation}
Let $\mathbf{x}_2(k)$ be the signal transmitted by Bob in phase 2 (echoing phase),
and the corresponding signals received by Alice and Eve are
respectively $\mathbf{y}_1(k)$ and $\mathbf{z}_2(k)$:
\begin{equation}\label{phase2}
    \begin{aligned}
        \mathbf{y}_1(k)&=\mathbf{H}_1\mathbf{x}_2(k)+\mathbf{w}_1(k),\\
        \mathbf{z}_2(k)&=\mathbf{G}_2\mathbf{x}_2(k)+\mathbf{v}_2(k).
    \end{aligned}
\end{equation}
Here $\mathbf{H}_2\in\mathbb{C}^{n_2\times n_1}$, $\mathbf{H}_1\in\mathbb{C}^{n_1\times n_2}$, $\mathbf{G}_1\in\mathbb{C}^{l_1\times n_1}$, $\mathbf{G}_2\in\mathbb{C}^{l_2\times n_2}$ are the corresponding channel matrices.
All noise vectors $\mathbf{w}_1(k)$, $\mathbf{w}_2(k)$, $\mathbf{v}_1(k)$ and $\mathbf{v}_2(k)$ are assumed to be $\mathcal{CN}(0,\mathbf{I})$. We do not assume full-duplex transceivers.

As a generalization of the round-trip precoding for Gaussian signaling used in STEEP shown in \cite{Hua2025}, we now let
\begin{equation}\label{precoders}
    \begin{aligned}
        \mathbf{x}_1(k)&=\mathbf{W}_1\mathbf{s}_1(k),\\
        \mathbf{x}_2(k)&=\mathbf{W}_y\mathbf{y}_2(k)+\mathbf{W}_s\mathbf{s}_2(k),
    \end{aligned}
\end{equation}
where $\mathbf{s}_1(k)$ is the probing signal,  $\mathbf{s}_2(k)$ is the secret-message dependent vector, and both are assumed to be $\mathcal{CN}(0,\mathbf{I})$.
Here $\mathbf{W}_1\in\mathbb{C}^{n_1\times n_1}$, $\mathbf{W}_y\in\mathbb{C}^{n_2\times n_2}$ and $\mathbf{W}_s\in\mathbb{C}^{n_2\times n_2}$ are the round-trip precoding matrices.

Following a similar analysis as shown in \cite{Hua2025}, we can show that an achievable secrecy rate of the above MIMO STEEP scheme in terms of $\mathbf{s}_2(k)$ from Bob to Alice, in bits per channel use or bits/s/Hz, is $\frac{1}{2}R_s^+=\max(0,\frac{1}{2}R_s)$ with
\begin{equation}\label{Rs}
    R_s=C_U-C_E
\end{equation}
where $C_U$ and $C_E$ are respectively the effective capacities at Alice and Eve to receive the information in $\mathbf{s}_2(k)$, i.e.,
\begin{equation}\label{eq:CU}
    \begin{aligned}
        &C_U=\mathbb{I}(\mathbf{s}_2(k);\mathbf{x}_1(k),\mathbf{y}_1(k))=\log\frac{1}{\lvert{\mathbf{K}_{\Delta s_2\vert x_1,y_1}\rvert}}\\
        &=\log\frac{\lvert\mathbf{H}_1\mathbf{W}_y\mathbf{W}_y^H\mathbf{H}_1^H+\mathbf{H}_1
        \mathbf{W}_s\mathbf{W}_s^H\mathbf{H}_1^H+\mathbf{I}\rvert}
        {\lvert\mathbf{H}_1\mathbf{W}_y\mathbf{W}_y^H\mathbf{H}_1^H+\mathbf{I}\rvert},
        \end{aligned}
        \end{equation}
        \begin{equation}\label{eq:CE}
         \begin{aligned}
        &C_E=\mathbb{I}(\mathbf{s}_2(k);\mathbf{z}_1(k),\mathbf{z}_2(k))=\log\frac{1}{\lvert{\mathbf{K}_{\Delta s_2\vert z_1,z_2}}\rvert}\\
        &=\log\frac{\lvert{\mathbf{G}_2\mathbf{W}_y\mathbf{K}\mathbf{W}_y^H\mathbf{G}_2^H+
        \mathbf{G}_2\mathbf{W}_s\mathbf{W}_s^H{\mathbf{G}_2^H}+\mathbf{I}}\rvert}
        {\lvert{\mathbf{G}_2\mathbf{W}_y\mathbf{K}\mathbf{W}_y^H\mathbf{G}_2^H+\mathbf{I}}\rvert},
    \end{aligned}
\end{equation}
with $|\cdot|\doteq\det(\cdot)$ and
\begin{equation}\label{K}
    \begin{aligned}
        &\mathbf{K}\doteq\mathbf{K}_{\Delta y_2\vert z_1}
        =\mathbf{H}_2\mathbf{W}_1\mathbf{W}_1^H\mathbf{H}_2^H+\mathbf{I}\\
        &\,\,-\mathbf{H}_2\mathbf{W}_1\mathbf{W}_1^H\mathbf{G}_1^H(\mathbf{G}_1\mathbf{W}_1
        \mathbf{W}_1^H\mathbf{G}_1^H+\mathbf{I})^{-1}\\
        &\,\,\cdot\mathbf{G}_1\mathbf{W}_1\mathbf{W}_1^H\mathbf{H}_2^H.
    \end{aligned}
\end{equation}
Here $\mathbf{K}_{\Delta y_2\vert z_1}$ (for example) denotes the MSE matrix of the MMSE estimate of $\mathbf{y}_2$ from $\mathbf{z}_1$.

 Assuming $n_1\geq n_2$, the precoders considered in \cite{Hua2025} are equivalent to
\begin{equation}\label{BL}
  \left \{\begin{array}{c}
            \mathbf{W}_1 = \sqrt{\frac{p_1}{n_1}}\mathbf{I}_{n_1}, \\
            \mathbf{W}_y=\sqrt{a}\sqrt{\frac{p_1}{n_1}}\boldsymbol{\Sigma}_2(
            \frac{p_1}{n_1}\boldsymbol{\Sigma}_2^2+\mathbf{I}_{n_2})^{-1}
            \mathbf{U}_2^H,\\
            \mathbf{W}_s = \sqrt{a}\mathbf{I}_{n_2},
          \end{array}
  \right .
\end{equation}
where $\boldsymbol{\Sigma}_2$ and $\mathbf{U}_2$ are from the (thin) SVD $\mathbf{H}_2=\mathbf{U}_2\boldsymbol{\Sigma}_2\mathbf{V}_2^H$, and $a$ is such that
\begin{equation}\label{}
  a\left (\texttt{Tr}\left (\frac{p_1}{n_1}\boldsymbol{\Sigma}_2^2
  \left (\frac{p_1}{n_1}\boldsymbol{\Sigma}_2^2+\mathbf{I}_{n_2}\right )^{-1}\right )+n_2\right )=p_2.
\end{equation}
The above precoders depend on $\mathbf{H}_2$ but are invariant to  $(\mathbf{H}_1,\mathbf{G}_1,\mathbf{G}_2)$, which will be referred as a baseline choice. It is shown in \cite{Hua2025} that, even using the baseline precoders, $R_s$ can be made positive by using sufficiently large $p_1$ and $\frac{p_2}{p_1}$ for virtually any given $(\mathbf{H}_1,\mathbf{H}_2,\mathbf{G}_1,\mathbf{G}_2)$. This property of MIMO STEEP is in great contrast to the conventional (single-trip) MIMO WTC schemes, the latter of which cannot produce a positive secrecy rate for a large channel space of $(\mathbf{H}_1,\mathbf{H}_2,\mathbf{G}_1,\mathbf{G}_2)$ if $l_1\geq n_1$ and $l_2\geq n_2$ \cite{KhistiWornell2010}, \cite{OggierHassibi2011}.

In this paper, we explore the performance of MIMO STEEP under optimized precoders. Namely, we are interested in the following problem for MIMO STEEP (which may also be viewed as MIMO RT-WTC problem):
\begin{subequations}\label{eq:optimization}
    \begin{align}
        &\max_{\mathbf{W}_1,\mathbf{W}_y,\mathbf{W}_s}\quad R_s,\label{}\\
        \operatorname{s.t.}\quad &\operatorname{Tr}(\mathbf{W}_1\mathbf{W}_1^H)\doteq p_1^*\leq p_1,\label{cons1}\\
        \operatorname{Tr}(\mathbf{W}_y&\mathbf{K}_{y_2}\mathbf{W}_y^H)+
        \operatorname{Tr}(\mathbf{W}_s\mathbf{W}_s^H)\doteq p_2^*\leq p_2,\label{cons2}
    \end{align}
\end{subequations}
where $\mathbf{K}_{y_2}=\mathbf{H}_2\mathbf{W}_1\mathbf{W}_1^H\mathbf{H}_2^H+\mathbf{I}$, and $p_1$ and $p_2$ are the allocated upper bounds on powers (or ``allowed'' powers) at Alice and Bob respectively while $p_1^*$ and $p_2^*$  denote the ``actual'' powers consumed by  Alice and Bob.

\section{Optimization}
To solve the problem \eqref{eq:optimization}, we first notice that with fixed $\mathbf{W}_1$, the constraint \eqref{cons2} is convex while \eqref{cons1} is not needed. With fixed $\mathbf{W}_2\doteq(\mathbf{W}_y,\mathbf{W}_s)$, both \eqref{cons1} and \eqref{cons2} are convex constaints. This makes AO a natural choice. So we propose to optimize $\mathbf{W}_1$ and $\mathbf{W}_2$ alternately.

With fixed $\mathbf{W}_1$, $R_s$ remains non-concave of $\mathbf{W}_2$. But its structure allows the application of WMMSE and SCA as shown later. With fixed $\mathbf{W}_2$, the structure of $R_s$ in terms of $\mathbf{W}_1$ does not allow the use of WMMSE, and in this case we will apply the more general principle PGA.

\subsection{Optimizing $(\mathbf{W}_y,\mathbf{W}_s)$ given $\mathbf{W}_1$}

With given $\mathbf{W}_1$, we can write
\begin{equation}\label{}
    \tilde{\mathbf{H}}_2 \doteq \mathbf{H}_2 \mathbf{W}_1, \quad \tilde{\mathbf{G}}_1 \doteq \mathbf{G}_1 \mathbf{W}_1,
\end{equation}
and $\mathbf{K}$ and $\mathbf{K}_{y_2}$ are constant matrices:
\begin{equation}\label{}
    \begin{aligned}
        &\mathbf{K}=  \tilde{\mathbf{H}}_2\tilde{\mathbf{H}}_2^H+\mathbf{I}-\tilde{\mathbf{H}}_2\tilde{\mathbf{G}}_1^H(\tilde{\mathbf{G}}_1 \tilde{\mathbf{G}}_1^H+\mathbf{I})^{-1}\tilde{\mathbf{G}}_1\tilde{\mathbf{H}}_2^H,\\
        &\mathbf{K}_{y_2}=  \tilde{\mathbf{H}}_2\tilde{\mathbf{H}}_2^H+\mathbf{I}.
    \end{aligned}
\end{equation}
In this case, \eqref{eq:optimization} reduces to
\begin{equation}\label{newopti}
    \begin{aligned}
       & \max_{(\mathbf{W}_y,\mathbf{W}_s)\in \mathcal{C}_2}\quad R_s, \quad \mathcal{C}_2=\{(\mathbf{W}_y,\mathbf{W}_s):\eqref{cons2}\}.
    \end{aligned}
\end{equation}
The objective function of the problem \eqref{newopti} can be written as:
\begin{equation}\label{goodbad}
    \begin{aligned}
        &R_s=\underbrace{[\log|\mathbf{J}_1|+\log|\mathbf{J}_4|]}_{T_{good}}
        \underbrace{-[\log|\mathbf{J}_2|+\log|\mathbf{J}_3|]}_{T_{bad}},
    \end{aligned}
\end{equation}
where
\begin{equation}\label{Js}
    \begin{aligned}
      \mathbf{J}_1&=\mathbf{H}_1\mathbf{W}_y\mathbf{W}_y^H\mathbf{H}_1^H
      +\mathbf{H}_1\mathbf{W}_s\mathbf{W}_s^H\mathbf{H}_1^H+\mathbf{I},\\
      \mathbf{J}_2&=\mathbf{H}_1\mathbf{W}_y\mathbf{W}_y^H\mathbf{H}_1^H+\mathbf{I},\\
      \mathbf{J}_3&=\mathbf{G}_2\mathbf{W}_y\mathbf{K}\mathbf{W}_y^H\mathbf{G}_2^H
      +\mathbf{G}_2\mathbf{W}_s\mathbf{W}_s^H{\mathbf{G}_2^H}+\mathbf{I},\\
      \mathbf{J}_4&=\mathbf{G}_2\mathbf{W}_y\mathbf{K}\mathbf{W}_y^H\mathbf{G}_2^H+\mathbf{I}.
    \end{aligned}
\end{equation}
Here $\log|\mathbf{J}|$ is a concave function of $\mathbf{J}$. But $R_s$ is a not concave function of $\mathbf{W}_y$ and $\mathbf{W}_s$. And the problem \eqref{newopti} is still non-convex. To tackle this, we next apply the principles of SCA and WMMSE.

\subsubsection{Applying SCA}
We now apply SCA to the last two terms in \eqref{goodbad}.
Since $-\log|\mathbf{J}|$ is a convex function of $\mathbf{J}$, it is lower bounded by its ``surrogate'' as follows:
\begin{equation}\label{surrogate}
    \begin{aligned}
        &-\log|\mathbf{J}|\ge-\log|\mathbf{J}^{(k)}|\\
        &\,\,- \operatorname{Tr}\bigg(\big(\nabla_{\mathbf{J}} \log|\mathbf{J}^{(k)}|\big)^H(\mathbf{J}-\mathbf{J}^{(k)})\bigg)\\
        &=-\log|\mathbf{J}^{(k)}|-\operatorname{Tr}\big({\mathbf{J}^{(k)}}^{-1}
        (\mathbf{J}-\mathbf{J}^{(k)})\big)\\
        &=\underbrace{-\log|\mathbf{J}^{(k)}|+\operatorname{Tr}(\mathbf{I})}_{\text{const}}-
        \operatorname{Tr}({\mathbf{J}^{(k)}}^{-1}\mathbf{J})\doteq T^{surro},
    \end{aligned}
\end{equation}
where $\mathbf{J}^{(k)}$ denotes the value of the matrix $\mathbf{J}$ evaluated at the $k$th iteration of $\mathbf{W}_y$ and $\mathbf{W}_s$, and we have used  $\nabla_\mathbf{X}\log|\mathbf{X}|=\mathbf{X}^{-1}$ \cite{Boydconvex}.

Applying \eqref{surrogate} to $-\log|\mathbf{J}_3|$, we have
\begin{equation}\label{}
    \begin{aligned}
        &T_3^{surro}=
        \text{const}-\operatorname{Tr}(\mathbf{W}_y^H\mathbf{G}_2^H{\mathbf{J}_3^{(k)}}^{-1}
        \mathbf{G}_2\mathbf{W}_y\mathbf{K})\\
        &\,\,-\operatorname{Tr}(\mathbf{W}_s^H\mathbf{G}_2^H{\mathbf{J}_3^{(k)}}^{-1}\mathbf{G}_2
        \mathbf{W}_s).
    \end{aligned}
\end{equation}
Similarly, for $-\log|\mathbf{J}_2|$, we have
\begin{equation}\label{}
    T_2^{surro}=\text{const}-\operatorname{Tr}(\mathbf{W}_y^H\mathbf{H}_1^H{\mathbf{J}_2^{(k)}}^{-1}
    \mathbf{H}_1\mathbf{W}_y).
\end{equation}

Therefore, with the $k$th iteration of $\mathbf{W}_y$ and $\mathbf{W}_s$, the last two terms (or $T_{bad}$) in \eqref{goodbad} can be replaced by
\begin{equation}\label{Jsurro}
    \begin{aligned}
        T^{surro}=&
        -\operatorname{Tr}(\mathbf{W}_y^H\mathbf{G}_2^H{\mathbf{J}_3^{(k)}}^{-1}\mathbf{G}_2
        \mathbf{W}_y\mathbf{K})\\
        &-\operatorname{Tr}(\mathbf{W}_s^H\mathbf{G}_2^H{\mathbf{J}_3^{(k)}}^{-1}
        \mathbf{G}_2\mathbf{W}_s)\\
        &-\operatorname{Tr}(\mathbf{W}_y^H\mathbf{H}_1^H{\mathbf{J}_2^{(k)}}^{-1}
        \mathbf{H}_1\mathbf{W}_y),
    \end{aligned}
\end{equation}
where all constant terms have been dropped.

\subsubsection{Applying WMMSE}
We now apply the principle of WMMSE to the first two terms in \eqref{goodbad}.
For $\mathbf{J}_1$, we define
\begin{equation}\label{defV1}
    \mathbf{V}_1 \doteq [\mathbf{W}_y\ \ \mathbf{W}_s]\in\mathbb{C}^{n_2\times (2n_2)}.
\end{equation}
Then
\begin{equation}\label{J1V1}
    \mathbf{J}_1 = \mathbf{I}+\mathbf{H}_1\mathbf{V}_1\mathbf{V}_1^H\mathbf{H}_1^H.
\end{equation}
For $\mathbf{J}_4$, we let $\mathbf{K}=\mathbf{L}\mathbf{L}^H$ and define
\begin{equation}\label{defV4}
    \mathbf{V}_4 \doteq \mathbf{W}_y\mathbf{L}\in\mathbb{C}^{n_2\times n_2},
\end{equation}
so that
\begin{equation}\label{J4V4}
    \mathbf{J}_4 = \mathbf{I}+\mathbf{G}_2\mathbf{V}_4\mathbf{V}_4^H\mathbf{G}_2^H.
\end{equation}

Define $\mathbf{y}=\mathbf{H}\mathbf{V}\mathbf{s}+\mathbf{n}$ with $\mathbf{s}\sim\mathcal{CN}(\mathbf{0},\mathbf{I})$ and $\mathbf{n}\sim\mathcal{CN}(\mathbf{0},\mathbf{I})$. A linear receiver $\hat{\mathbf{s}}=\mathbf{U}^H\mathbf{y}$ has the following MSE matrix
\begin{equation}\label{defMSE}
    \begin{aligned}
        &\mathbf{E}(\mathbf{U},\mathbf{V})
        \doteq \mathbb{E}[(\mathbf{s}-\mathbf{U}^H\mathbf{y})(\mathbf{s}-\mathbf{U}^H\mathbf{y})^H]\\
        &=\mathbf{I}-\mathbf{U}^H\mathbf{H}\mathbf{V}-(\mathbf{U}^H\mathbf{H}\mathbf{V})^H+
        \mathbf{U}^H(\mathbf{H}\mathbf{V}\mathbf{V}^H\mathbf{H}^H+\mathbf{I})\mathbf{U}.
    \end{aligned}
\end{equation}

Recall  the WMMSE identity \cite{Li2013},  \cite{Jose2011}:
\begin{align}\label{rateWMMSE}
&\log|\mathbf{I}+\mathbf{H}\mathbf{V}\mathbf{V}^H\mathbf{H}^H|\notag\\
&=\max_{\mathbf{U},\ \mathbf{S}\succ\mathbf{0}}
\ \log|\mathbf{S}|-\operatorname{Tr}(\mathbf{S}\mathbf{E}(\mathbf{U},\mathbf{V}))+d,
\end{align}
where $d$ is the number of columns of $\mathbf{V}$. Given $\mathbf{S}$ and $\mathbf{U}$, $-\operatorname{Tr}(\mathbf{S}\mathbf{E}(\mathbf{U},\mathbf{V}))$ is a concave quadratic function of $\mathbf{V}$.

Applying \eqref{rateWMMSE} to $\log|\mathbf{J}_1|$ and $\log|\mathbf{J}_4|$ yields
\begin{equation}\label{wmmseJ1}
    \begin{aligned}
        &\log|\mathbf{J}_1|\\
        &=\max_{\mathbf{U}_1,\mathbf{S}_1\succ\mathbf{0}}
        \log|\mathbf{S}_1|-\operatorname{Tr}(\mathbf{S}_1\mathbf{E}_1(\mathbf{U}_1,\mathbf{V}_1))+2n_2,\\
        &\log|\mathbf{J}_4|\\
        &=\max_{\mathbf{U}_4,\mathbf{S}_4\succ\mathbf{0}}
        \log|\mathbf{S}_4|-\operatorname{Tr}(\mathbf{S}_4\mathbf{E}_4(\mathbf{U}_4,\mathbf{V}_4))+n_2,
    \end{aligned}
\end{equation}
where $\mathbf{E}_1(\mathbf{U}_1,\mathbf{V}_1)$ and $\mathbf{E}_4(\mathbf{U}_4,\mathbf{V}_4)$ follow \eqref{defMSE}.

Given $(\mathbf{W}_y^{(k)},\mathbf{W}_s^{(k)})$, we have $\mathbf{V}_1^{(k)}=[\mathbf{W}_y^{(k)}\ \mathbf{W}_s^{(k)}]$ and $\mathbf{V}_4^{(k)}=\mathbf{W}_y^{(k)}\mathbf{L}$. And the corresponding optimal $\mathbf{U}_1$ and $\mathbf{U}_4$ are
\begin{equation}\label{closeformU}
    \begin{aligned}
        \mathbf{U}_1^{(k+1)}=(\mathbf{H}_1\mathbf{V}_1^{(k)}(\mathbf{V}_1^{(k)})^H\mathbf{H}_1^H+
        \mathbf{I})^{-1}\mathbf{H}_1\mathbf{V}_1^{(k)},\\
        \mathbf{U}_4^{(k+1)}=(\mathbf{G}_2\mathbf{V}_4^{(k)}(\mathbf{V}_4^{(k)})^H\mathbf{G}_2^H+
        \mathbf{I})^{-1}\mathbf{G}_2\mathbf{V}_4^{(k)}.
    \end{aligned}
\end{equation}
Let $\mathbf{E}_1^{(k+1)}\doteq\mathbf{E}_1(\mathbf{U}_1^{(k+1)},\mathbf{V}_1^{(k)})$ and $\mathbf{E}_4^{(k+1)}\doteq\mathbf{E}_4(\mathbf{U}_4^{(k+1)},\mathbf{V}_4^{(k)})$.
Then the optimal weights are
\begin{equation}\label{closeformS_new}
    \begin{aligned}
    \mathbf{S}_1^{(k+1)}=(\mathbf{E}_1^{(k+1)})^{-1},\\
    \mathbf{S}_4^{(k+1)}=(\mathbf{E}_4^{(k+1)})^{-1}.
    \end{aligned}
\end{equation}

\subsubsection{Combining SCA and WMMSE}
Combining the WMMSE representation for $T_{good}$ and the SCA lower bound for $T_{bad}$ in \eqref{goodbad}, $R_s$ with the $k$th update of  $(\mathbf{W}_y,\mathbf{W}_s)$ can be represented by
\begin{equation}\label{eq:gk}
    \begin{aligned}
        &g^{(k)}
        =-\operatorname{Tr}\big(\mathbf{S}_1^{(k+1)}\mathbf{E}_1(\mathbf{U}_1^{(k+1)},\mathbf{V}_1)\big)\\
        &-\operatorname{Tr}\big(\mathbf{S}_4^{(k+1)}\mathbf{E}_4(\mathbf{U}_4^{(k+1)},\mathbf{V}_4)\big)
        +T^{surro}(\mathbf{W}_y,\mathbf{W}_s)
    \end{aligned}
\end{equation}
which is a concave quadratic function of $(\mathbf{W}_y,\mathbf{W}_s)$. Here $\mathbf{V}_1=[\mathbf{W}_y\ \mathbf{W}_s]$ and $\mathbf{V}_4=\mathbf{W}_y\mathbf{L}$. Therefore, the problem \eqref{newopti} can be solved iteratively by solving the following convex problem:
\begin{equation}\label{final}
    \begin{aligned}
        \max_{\mathbf{W}_y,\mathbf{W}_s\in\mathcal{C}_2} & g^{(k)},
    \end{aligned}
\end{equation}
which maximizes a concave quadratic objective over a convex quadratic constraint set.

\subsection{Optimizing $\mathbf{W}_1$ given $(\mathbf{W}_y,\mathbf{W}_s)$}\label{sec:computing_W1}

With given $(\mathbf{W}_y,\mathbf{W}_s)$, \eqref{eq:optimization} reduces to
\begin{equation}\label{eq:W1_subprob}
    \max_{\mathbf{W}_1\in\mathcal{C}_1}\; R_s,
    \quad
    \mathcal{C}_1 \doteq \left\{\mathbf{W}_1:\eqref{cons1},\eqref{cons2}\right\}.
\end{equation}
Here $\mathcal{C}_1$ is convex but the dependency of $R_s$ on $\mathbf{W}_1$ (see \eqref{K}) does not allow the use of WMMSE. But we will use the PGA principle \cite{Bertsekasnlp} as follows:
\begin{equation}\label{eq:W1_PGA_update}
    \begin{aligned}
        \mathbf{W}_{1,\mathrm{tmp}}&\leftarrow \mathbf{W}_1 + \tau \nabla_{\mathbf{W}_1}R_s\!\left(\mathbf{W}_1,\mathbf{W}_y,\mathbf{W}_s\right)\\
        \mathbf{W}_1&\leftarrow\Pi_{\mathcal{C}_1}\!\left(\mathbf{W}_{1,\mathrm{tmp}}\right)
    \end{aligned}
\end{equation}
where $\tau$ is chosen by Armijo backtracking \cite{Armijo1966}, and $\Pi_{\mathcal{C}_1}(\cdot)$ denotes the projection onto $\mathcal{C}_1$.

By alternate optimization (AO) between  $(\mathbf{W}_y,\mathbf{W}_s)$ and $\mathbf{W}_1$, we can obtain locally optimal $(\mathbf{W}_1, \mathbf{W}_y,\mathbf{W}_s)$.

\section{Simulation}
To illustrate the results of the optimized secrecy rate of MIMO STEEP, we consider  a scenario where Eve is at the middle between Alice and Bob. The distance between Alice and Bob is normalized to be one, and the channel matrices between them are subject to Rician fading. Specifically, we assume
\begin{equation}\label{}
  \mathbf{H}_2 = \sqrt{K/(K+1)}\mathbf{a}_r\mathbf{a}_t^He^{j\theta_2}+\sqrt{1/(K+1)}\mathbf{N}_2
\end{equation}
where $K$ is the Rician factor, $\mathbf{N}_2$ consists of i.i.d. elements from $\mathcal{CN}(0,1)$, $\theta_2$ is a uniform random phase, and $\mathbf{a}_r$ and $\mathbf{a}_t$ are receive and transmit steering vectors of linear arrays. The choice of $\mathbf{H}_1$ is similar. But $\mathbf{G}_1$ and $\mathbf{G}_2$ are scaled up by the factor 2 (assuming pathloss exponent equal to 2) while they have the similar structures as $\mathbf{H}_1$ and $\mathbf{H}_2$. Also assume that all arrays are oriented toward each other (so that the $\mathbf{a}$-vectors consist of all ones). Also let $n_E=l_1=l_2$, $K=1$ (or 0dB), and recall $\mathbf{W}_2\doteq(\mathbf{W}_y,\mathbf{W}_s)$.

\begin{figure}[t]
%

    \subfloat[WTC vs AN vs BL. $n_E=8$, $(p_1,p_2)=(20,40)$dB.]{
        \includegraphics[width=0.45\linewidth]{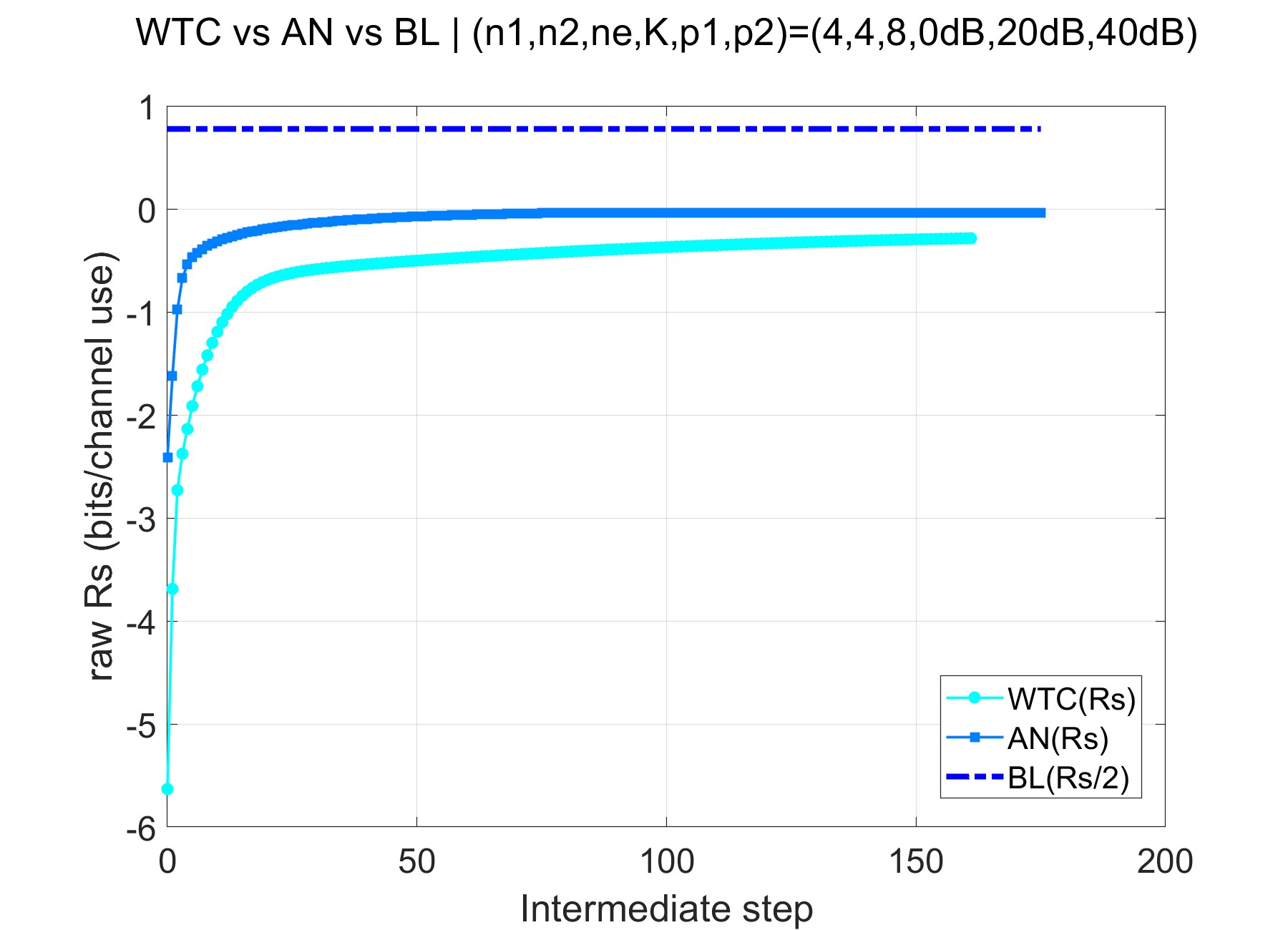}
        \label{fig:sub1}
    }\hfill
    \subfloat[Opt. STEEP vs BL. $n_E=6$, $(p_1,p_2)=(5,20)$dB.]{
        \includegraphics[width=0.45\linewidth]{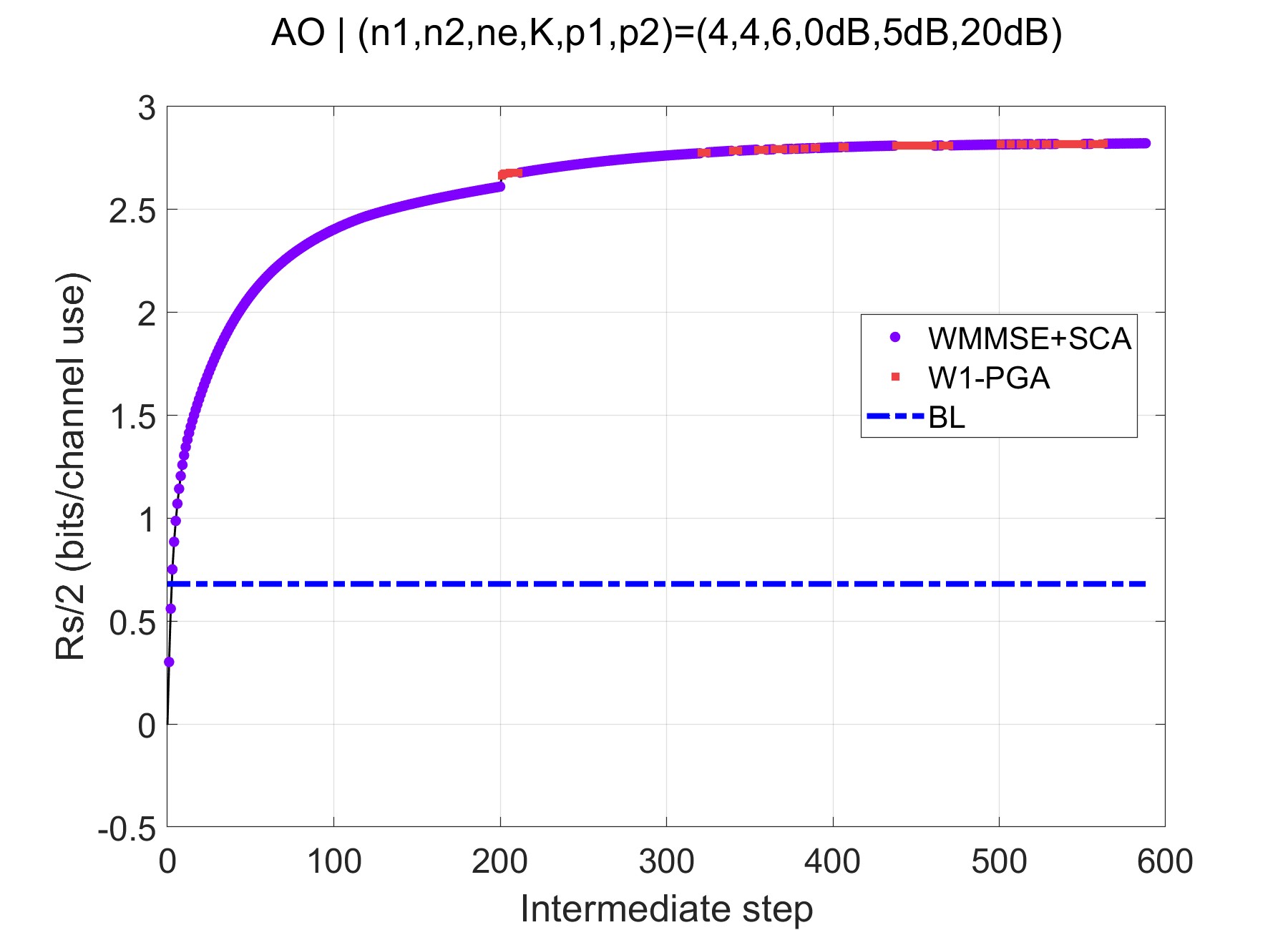}
        \label{fig:sub2}
    }

    \caption{Convergence patterns of $R_s$ using $n_1=n_2=4$.}
    \label{fig:1}
\end{figure}

Fig. \ref{fig:sub1} shows a case where the optimized $R_s$ of the classic ST-WTC scheme \cite{KhistiWornell2010}-\cite{OggierHassibi2011} (equivalently using $p_1=0$ here)  can only converge to zero while STEEP using the baseline (BL) precoders shown in \eqref{BL} and $(p_1,p_2)=(20,40)$dB yields a significant  $\frac{1}{2}R_s>0$. In this figure, WTC denotes the case for $\mathbf{W}_y=0$ and optimized $\mathbf{W}_s$, and AN (artificial noise) denotes the case  for optimized $(\mathbf{W}_y,\mathbf{W}_s)$. Both WTC and AN converge to the same secrecy rate (zero here). It is known that WTC and AN yield the same $R_s$ after optimization. All three curves in Fig. \ref{fig:sub1} are based on a common random realization of $(\mathbf{H}_1,\mathbf{H}_2,\mathbf{G}_1,\mathbf{G}_2)$.

Fig. \ref{fig:sub2} shows that $\frac{1}{2}R_s$ of STEEP using the optimized precoders is much higher than that using the baseline (BL) precoders in \eqref{BL}, where $(n_1,n_2,n_E)=(4,4,6)$ and $(p_1,p_2)=(5,20)$dB. This figure also shows the convergence behavior of $R_s$ when optimizing $\mathbf{W}_1$ and $\mathbf{W}_2$ alternately. The purple dots correspond to $\mathbf{W}_2$ being optimized while the red dots correspond to $\mathbf{W}_1$ being optimized.

\begin{figure}[t]
%

    \subfloat[$\frac{1}{2}R_s$ of Baseline STEEP]{
        \includegraphics[width=0.45\linewidth]{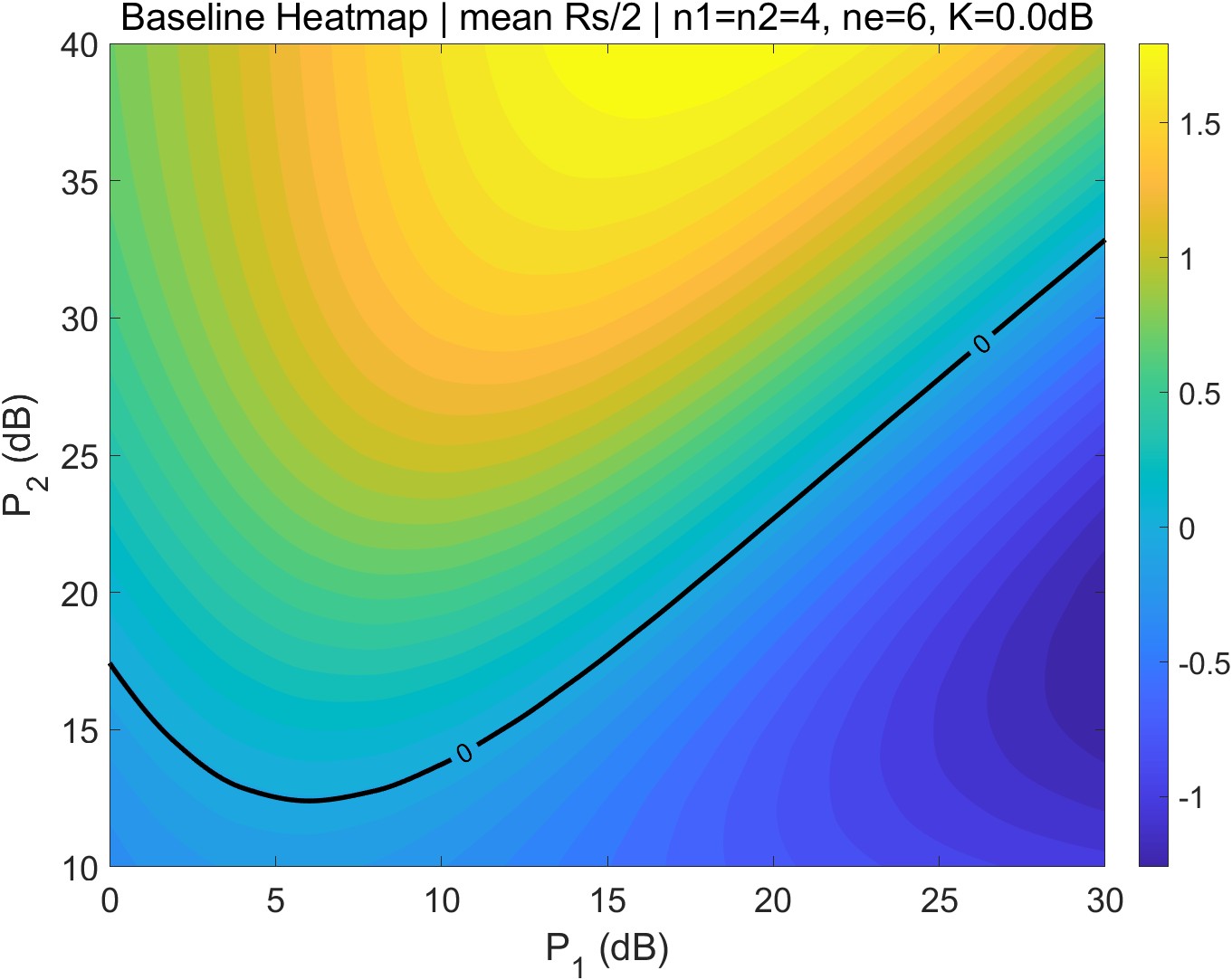}
        \label{fig:sub3}
    }\hfill
    \subfloat[$\frac{1}{2}R_s$ of Optimized STEEP]{
        \includegraphics[width=0.45\linewidth]{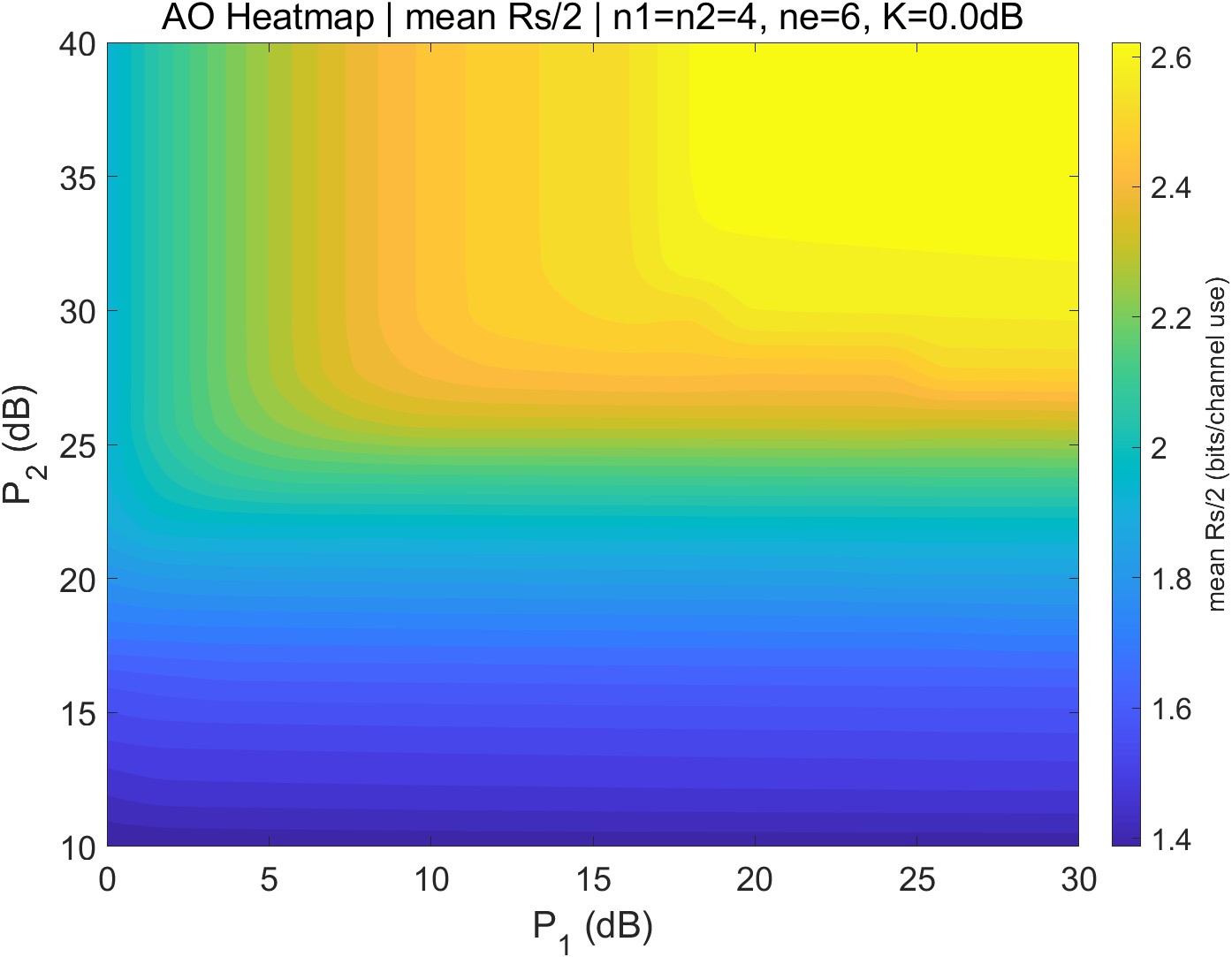}
        \label{fig:sub4}
    }

    \caption{Averaged heatmaps of $\frac{1}{2}R_s$ using $(n_1,n_2,n_E)=(4,4,6)$. For (a), the axes are the ``actual'' powers $(p_1^*,p_2^*)$. For (b), the axes are the ``allowed'' powers $(p_1,p_2)$.}
    \label{fig:2}
\end{figure}

Fig. \ref{fig:sub3} shows an averaged heatmap of $\frac{1}{2}R_s$ of the baseline STEEP versus $(p_1^*,p_2^*)$. For each $(p_1^*,p_2^*)$,  10 realizations of $(\mathbf{H}_1,\mathbf{H}_2,\mathbf{G}_1,\mathbf{G}_2)$ with $(n_1,n_2,n_E)=(4,4,6)$ were used. As expected, $R_s>0$ if $p_1^*$ and $\frac{p_2^*}{p_1^*}$ are large enough.

Fig. \ref{fig:sub4} shows a similar heatmap of the optimized STEEP versus $(p_1,p_2)$.
We see that $R_s$ saturates when $p_1$ and $p_2$ are beyond some values. Since $R_s$ typically increases with $p_2^*$, the optimal $p_2^*$ is $p_2$. But similar to Fig. \ref{fig:sub3}, the optimal value of $p_1^*$ increases with $p_2^*$ (due to ``strong'' Eve) but  is much (about 10-15dB) smaller than $p_2^*$. Fig. \ref{fig:sub4} unfortunately conveys little information about the corresponding values of $(p_1^*,p_2^*)$, which cannot be discussed further due to page limit.

\section{Conclusion}

For the first time, we have shown how to compute the optimal precoders for MIMO STEEP, which is a round-trip scheme important for delivering a positive secrecy rate even when Eve's channels are much stronger than users'. Among the important observations, we have found that, with the exact knowledge of Eve's channels, the secrecy rate of MIMO STEEP using optimized precoders can be much higher than that using the baseline precoders.
Future work needs to consider optimal precoder design with only limited knowledge of Eve such as her worst possible location and maximal possible number of antennas in a given application.


\newpage

\begin{thebibliography}{00}

%
%

 \bibitem{KhistiWornell2010}
 A. Khisti, G. W. Wornell,  ``Secure transmission with multiple antennas -- Part II: The MIMOME wiretap channel,'' IEEE Trans Inf Theory,
Vol. 56, pp.5515-5532, 2010.

 \bibitem{OggierHassibi2011}
 F. Oggier, B. Hassibi, ``The secrecy capacity of the MIMO wiretap channel'', IEEE
Trans Inf Theory, Vol. 57, pp.4961-4972, 2011.


\bibitem{RennaBloch2013}
F. Renna and R. Bloch, ``Semi-blind key-agreement over MIMO fading channel,'' IEEE Trans. Comm., Vol. 61, No. 2, Feb 2013.

\bibitem{HuaMaksud2024}
Y. Hua and A. Maksud, ``Secret-key capacity from MIMO channel probing,''  IEEE Wireless Communications Letters,  vol. 13, no. 5, pp.
 1434–1438, May 2024.

\bibitem{Hayashi2020}
M. Hayashi and A. Vazquez-Castro, ``Two-way physical layer security protocol for Gaussian channels,''
IEEE Transactions on Communications, Vol. 68, No. 5, pp. 3068-3078, 2020.

\bibitem{Hua2023}
Y. Hua, ``Generalized channel probing and generalized pre-processing for secret key generation,'' IEEE Transactions on Signal Processing, Vol. 71, pp. 1067-1082, April 2023.

\bibitem{Hua2025}
Y. Hua, ``On secret-message transmission by echoing encrypted probes,'' IEEE Transactions on Communications, vol. 73, no. 11, pp. 12053-12069, Nov. 2025.

 \bibitem{HuaICC2024}
 Y. Hua and M. S. Rahman, ``Unification of secret key generation and wiretap channel transmission,'' IEEE ICC, Denver, CO, June 2024.

 \bibitem{HuaLANMAN2024}
 Y. Hua, M. S. Rahman, and A. Swami, ``A method for low-latency secure multiple access,'' IEEE LANMAN, Boston, MA, July 2024 (Best Paper Award).


\bibitem{Rahman2024}
M. S. Rahman, and Y. Hua, ``Secure UAV communications by STEEP against full-duplex active eavesdropper,'' Asilomar Conference on Signals, Systems and Computers, Monterey, CA, Oct 2024.

\bibitem{Rafique2025}
M. Rafique and Y. Hua, ``Investigation of STEEP using phase-shift-keying signaling for secure communications,'' IEEE MILCOM, Los Angeles, CA, Oct. 2025.



\bibitem{Bloch2011}
M. Bloch and J. Barros, Physical-Layer Security, Cambridge University Press, 2011.

 \bibitem{Gamal2011}
 A. E. Gamal and Y.-H. Kim, Network Information Theory. Cambridge university press, 2011.


%


\bibitem{Shi2011}
Q. Shi, M. Razaviyayn, Z.-Q. Luo and C. He, ``An iteratively weighted MMSE approach to distributed sum-utility maximization for a MIMO interfering broadcast channel,'' IEEE Transactions on Signal Processing, vol. 59, no. 9, pp. 4331-4340, Sept. 2011.

\bibitem{Li2013}
Q. Li, M. Hong, H.-T. Wai, Y.-F. Liu, W.-K. Ma and Z.-Q. Luo, ``Transmit solutions for MIMO wiretap channels using alternating optimization,''  IEEE Journal on Selected Areas in Communications, vol. 31, no. 9, pp. 1714-1727, September 2013.

\bibitem{Mukherjee2022}
A. Mukherjee, V. Kumar, E. Jorswieck, B. Ottersten and L. -N. Tran, ``On the optimality of the stationary solution of secrecy rate maximization for MIMO wiretap channel,'' IEEE Wireless Communications Letters, vol. 11, no. 2, pp. 357-361, Feb. 2022.

\bibitem{Boydconvex}
S. Boyd and L. Vandenberghe, Convex optimization, Cambridge University Press, 2004.

\bibitem{Jose2011}
J. Jose, N. Prasad, M. Khojastepour and S. Rangarajan, ``On robust weighted-sum rate maximization in MIMO interference networks,'' 2011 IEEE International Conference on Communications (ICC), Kyoto, Japan, 2011, pp. 1-6.

\bibitem{Armijo1966}
L. Armijo, ``Minimization of functions having Lipschitz continuous first partial derivatives,'' Pacific Journal of mathematics 16.1 (1966): 1-3.

\bibitem{Bertsekasnlp}
D. P. Bertsekas,  ``Nonlinear programming,'' Journal of the Operational Research Society 48.3 (1997): 334-334.

\end{thebibliography}
\end{document}